\title{Testing and validating \textit{AnTraGoS} algorithms
	with impact beating spatters}
\author{
        Francesco Camana, Ph.D.\footnote{Italian State Police (\emph{Polizia di Stato}), Interregional forensic science office (\emph{G.I.P.S.}), Piazzetta Palatucci, 5 - 35123 Padova (ITALY)}, Massimiliano Gori, MD*, \\
        Luca De Rosa* and Roberto Mangione*
    }

\date{\today}

\documentclass[12pt]{article}
\usepackage{graphicx}
\usepackage{epstopdf}
\usepackage{amsfonts}
\usepackage{verbatim}
\usepackage[table]{xcolor}

\begin{document}
\maketitle

\begin{abstract}
The reconstruction of the area of origin of spatter patterns is usually a fundamental step to the determination of the area of the crime scene where the victim was wounded. In this field, for almost a decade, the italian \textit{Polizia di Stato} has employed \textit{AnTraGoS}, a forensic software which implements a probabilistic approach to identify the area where the horizontal projections of the trajectories of a set of blood drops converge (area of convergence) and to estimate the height of origin. In this paper we summarize a series of tests performed on a published dataset of spatter patterns, whose results confirm the validity of \textit{AnTraGoS} and of its algorithms. As a side result, some useful suggestions are derived, concerning the determination of the height of origin, within a statistical and fluid dynamic approach.
\end{abstract}

\begin{scriptsize}
\begin{center}
\end{center}
\end{scriptsize}

\section{Introduction}\label{Intro}
One distinguishing feature of Bloodstain Pattern Analysis (BPA) is the application of geometrical and physical concepts to the analysis of the bloodstains detected at a crime scene. Among other results, the methods and the outcomes of BPA may suggest how to group stains with similar origin, how to spot the location of the source of the projections, how to describe the temporal sequence of the occurred events. The reconstruction of the criminal dynamics and the verification of the testimonies at court often profit of these results: in particular, some key information about the absolute and relative position of the offender and of the victim during the different phases of the dynamics can also be argued and assessed \cite{BPD,MD1,BKNAST}.

Recent developements in this field effectively concern procedures and techniques for determining the area of convergence and the area of origin\footnote{According to the current terminology, the area of convergence is \emph{the space in two dimensions to which the directionalities of spatter stains can be retraced to determine the position of the spatter producing event} while the area of origin is \emph{the space in three dimensions to which the trajectories of spatter can be utilized to determine the location of the spatter producing event}.} of projected blood from single static sources \cite{A1}. These areas, which are commonly indicative of the position of the victim and of the offender at the time when the woundings were generated, can only be argued \emph{a posteriori}, by mean of a detailed analysis of the bloodstain pattern.

To this scope, different techniques can be used, ranging from the method of the strings (or trigonometric method), which assumes the trajectories are straight lines \cite{Par, Car}, to more complex fluid dynamics models \cite{Var, A3, Kaba, KD, RSL}, which can be implemented only by specifically coded algorithms. The method of the strings is particularly semplified, and reveals its weakness in neglecting both the gravity and the drag force, producing in this way a satisfactory agreement only at short distance and for fast, upward moving blood drops \cite{CIF}.

For more than a decade, the Italian \emph{Polizia di Stato} (State Police) has approached most of its BPA caseworks by employing a forensic software named \textit{AnTraGoS} (from the italian acronym of Analysis of the Trajectories of the Drops of Blood)\footnote{The program \emph{AnTraGoS} is a project designed and realized by Francesco Camana. Since 2008 \textit{AnTraGoS} is acknowledged and used by the Forensic Science Service of the Italian \emph{Polizia di Stato}.}. \textit{AnTraGoS} is a program which allows the calculus of the angles of impact of the blood drops in 3D, the estimation of the uncertainties, the numerical analysis of single trajectories (including gravity and air resistance), the definition of the area of convergence, the estimation of the area of origin by mean of the statistical analysis of the velocities of the drops, calculated at different probe-heights.

In detail, the method for the calculation of the area of convergence, by mean of a probabilistic approach, has been also published by one of us in 2013 \cite{CF}, and it has connections with works of other researchers  \cite{A1,CH}.

\textit{AnTraGoS} simply applies trigonometry, standard error analysis and classical mechanics. Direct tests have shown that the implemented algorithms and formulas are correct and produce consistent results. Nevertheless, with this work, we want to prove how and to what extent the results produced by \textit{AnTraGoS} are consistent with real experimental conditions. In this way, we also ascertain the applicability of the program for forensic cases and its scientific validity at court.

To do so we profit of a recently published data set of experimental impact spatter patterns, where the points of origin of the single projections is known. For every test, starting from the high-resolution image of the pattern and measuring the position of the single stains, we use \textit{AnTraGoS} to determine the area of convergence and to estimate geometrical and physical limits to the area of origin. The comparison between the results of the software and the real experimental conditions can delineate the potential field of validity of the algorithms and confirm the applicability of the software itself.

\section{Materials and methods}\label{ID}
\subsection{Materials}
The data set \cite{A2} consists in 61 swine fresh blood impact beating spatters, deposited on poster board sheets, scanned at 600 dots per inch. The image files of the patterns can be downloaded from the following website: \emph{https://www.ncbi.nlm.nih.gov/pmc/articles/PMC5996142/}. The images of the patterns are analysed both with \emph{Gimp} (GNU Image Manipulation Program v. 2.10.8) and \emph{Adobe Photoshop} v. 21.0.2. The version 2.4.2 of the software \textit{AnTraGoS} (released in 2014 and currently distributed) is used for the validation tests. All the cited programs are run on HP or DELL graphics workstations, with \emph{Windows 10} operative system. 

\subsection{Methods}

The procedure for the analysis is standard and follows this scheme: (a) stains selection \cite{Ill}; (b) measurement of the position of the stains; (c) measurement of the size and of the directionality of the stains \cite{Joris}; (d) calculation of the angles of impact \cite{BPD, Car} and of their relative uncertainties \cite{CF,Pa,WPDBR,R}; (e) detection of the area of convergence \cite{CF}; (f) reconstruction of the blood-drop trajectories and identification of the area of origin \cite{A1,K}.
More in detail:
\begin{itemize}
	\item[(a)] \textbf{stains selection}: 10 to 12 stains are selected from each pattern, paying attention to choose well-formed elliptical stains, uniformly distributed at different heights and horizontal positions, formed both from ascending and descending drops (it is a favourable property of \textit{AnTraGoS} the possibility to mix upward and downward moving drops together, following the procedure detailed in \cite{CF}).
	\item[(b)] \textbf{measurement of the position of the stains}: the position and the cartesian coordinates of the stains are estimated through an image analysis software, given the resolution of the image of the pattern and given the coordinates of the lower left corner of the poster board. The center of the ellipse of the stain is used to calculate this position and, given a certain overall uncertainty in the measurement process, the position of deposit is recorded with an estimated error of $\pm 1$ cm.
	\item[(c)] \textbf{measurement of the size and of the directionality of the stains}: the estimation of the minimum and maximum values of the dimensions of the axes of the stain is performed in the image analysis software, considering the color gradient of the pixels of the image near the border of the elliptical shape of the stain. Only the \textit{upper} part of the ellipse is considered. Analogously, the directionality of the stain with respect to the vertical line is estimated, ranging from a minimum and a maximum value, considering different reasonable approximations of the symmetry axis of the stain. A typical uncertainty for this measure is within the interval from $\pm 2^o$ to $\pm4^o$.
	\item[(d)] \textbf{calculation of the angles of impact and of their uncertainty} (see \cite{CF} for details and conventions): the angle $\alpha$ (and its uncertainty $\delta \alpha$) between the tangent to the trajectory at the point of impact and the surface of deposit is calculated in \textit{AnTraGoS}, by applying the following expressions:
	\begin{equation}
	\alpha=\arcsin{\left(d\over a\right)}
	\label{eq_ang_impatto}
	\end{equation}
	\begin{eqnarray}
	\delta \alpha=\sqrt {{{d^2}\over{a^2{\left(a^2-d^2\right)}}} \left(\delta a\right)^2+{{1}\over{\left(a^2-d^2\right)}} \left(\delta d\right)^2}.
	\label{eq_err_angimpatto}
	\end{eqnarray}
	where $d$ and $a$ are the best estimates of the dimensions of the minor and major axes of the stain and $\delta d$ and $\delta a$ their respective uncertainty.
	
	More relevantly, for stains deposited on vertical surfaces, the projection of the angle of impact onto the horizontal plane generates an angle  $\gamma$ (with an uncertainty $\delta \gamma$) with the surface itself (clockwise measured):
	\begin{equation}
	\gamma=\arctan {{{d\over a }}\over{ \sin \phi \sqrt{1-{{d^2}\over{a^2}}}}}
	\label{eq_ang_gamma}
	\end{equation}
	and
	\begin{eqnarray}
	\delta \gamma&=&\sqrt {{{a^4\sin^2\phi}\over{\left(a^2-d^2\right){\left[\sin^2\phi \left(a^2-d^2\right)+d^2\right]}^2}}\cdot}\nonumber\\
	&&\overline{\cdot\left[{{d^2}\over{a^2}}\left(\delta a\right)^2+\left(\delta d\right)^2+{d^2 \cos^2 \phi \over {a^4\sin^2 \phi}} {\left(a^2-d^2\right)}^2
		\left(\delta \phi \right)^2\right]}\qquad\qquad
	\label{eq_err_gamma}
	\end{eqnarray}
	
	where $\phi$ is the angle of the major axis, with respect to the vertical line, and $\delta \phi$ its uncertainty.
	
	\item[(e)] \textbf{detection of the area of convergence}: the identification of the area of convergence is achieved by employing the statistical method described in \cite{CF}. We specify here that the calculation of the PDF (probablity density function) is performed by \textit{AnTraGoS} on a grid of 1, 3 or 5 cm spacing and the normalization is calculated by considering a 100\% value of probability on a 2 m by 2 m square centered on the point of maximum probablility of convergence.
	This last choice is particularly required to maintain the integrals finite in cases where the angles of impact $\gamma$ are all similar and then, in principle, there is no convergence at all.
	\item[(f)] \textbf{reconstruction of the blood-drop trajectories and identification of the area of origin}: this step is the more complex and necessarily requires a fluid dynamic approach. The idea is however very simple, and follows in \textit{AnTraGoS} the procedure explained in \cite{A1}, with some variations and additional hypotheses. First of all, we notice that the trajectory of the drop, backward in time from the point of impact, need to be stopped at some time or some distance, physically compatible with the scene or the speed of the drop itself. In agreement with the idea that we want to determine the heigth of the point of origin, which is a point vertically aligned with the area of convergence, we decide to stop the calculation of the trajectory at a \emph{probe} point over the area determined in the step (e). The chosen probe point is the point of maximum probability of convergence. In a few words, we assume that the trajectory passes over the probe point (the assumption is in agreement with the definition of area of convergence) and, for every height in a presumptive interval (AnTraGos can test intervals of 120 cm, with 10 cm spacing, or intervals of 40 cm, with 2 cm spacing)  we calculate the velocity that the drop should have to impact properly (with the right angle of impact) onto the surface of deposit. In this way, a compared analysis of the velocities at different heights above the probe point can be made. These different velocities, relative to different drops and stains, can be \emph{e.g.} used to calculate averages, standard deviations and extreme values. Fluid dynamical and statistical considerations are then used to restrict the range of possible heights: for example heights which imply high velocities for drops whose size is incompatible with high speeds are excluded while heights which minimize averages and standard deviations are favoured (recollecting that the projections derive from a single impact event). More on this will be detailed in Par.\ref{results}.
\end{itemize}

A total of 12+6=18 sets of stains were selected in the overall data set: all the 12 produced by the cylinder rig, denominated \textbf{C{\scriptsize \#}}, and 6, randomly chosen, produced by the hockey puck rig, denominated \textbf{HP\_{\scriptsize \#}}. Among these 6, the set nr. \emph{HP\_63} is peculiar, in that the pattern is formed by two distint projections and origins: the results for this pattern are splitted into two parts, as if belonging to different sets.

Among the useful stains present in every pattern, regular in shape and definite in their edge characteristics, 10 to 12 of them have been selected. This number has proven to be sufficient for a statistical analysis and repeated tests have demonstrated that larger numbers of stains do not generate significant changes in the results, with the only unwanted result of linearily increasing the time needed for the analysis. 

The selection in the pattern is made according to the idea that, lying the stains on a single vertical deposit surface, a good convergence can be obtained only if some selected stains are located in proximity of the left and right border of the surface, in order to differentiate as much as possible the orientation of the horizontal projections of impact angles. A certain uniform distribution on the vertical axis is also preferred during the selection process. A balanced choice of ascending and descending drops is made, adequate to the proportion of upward and downward oriented stains on the impact surface.

\section{Results}\label{results}
We split the analysis of the results into two parts: in the first we describe the outputs of the module for the determination of the \textit{Area of Convergence}, in the second we discuss the results of the physical analysis which indicates how to determine the height and the \textit{Area of Origin}.\newline For the first part we refer to \cite{CF}, while for the area of origin we refer to \cite{A1} and to the ideas summarized below and in Appendix 1. In this last case, the equations of motion for the calculation of the trajectories and of the velocities are solved with \textit{AnTraGoS} by mean of a second order Runge-Kutta algorithm (details to be published). Air friction and gravity are obviously included in the method. \newline The coordinate system in \emph{AnTraGoS} is different from that used by Attinger \textit{et al.} in \cite{A2}. To make the results more easily understandable, we use a single system, by converting the coordinates in \cite{A2}. Here, as in \emph{AnTraGoS}, the vertical coordinate is $y$, while the horizontal plane is the $xz$ plane. The conversion is therefore as follows:
\\

\begin{minipage}{\textwidth} \includegraphics[width=\linewidth]{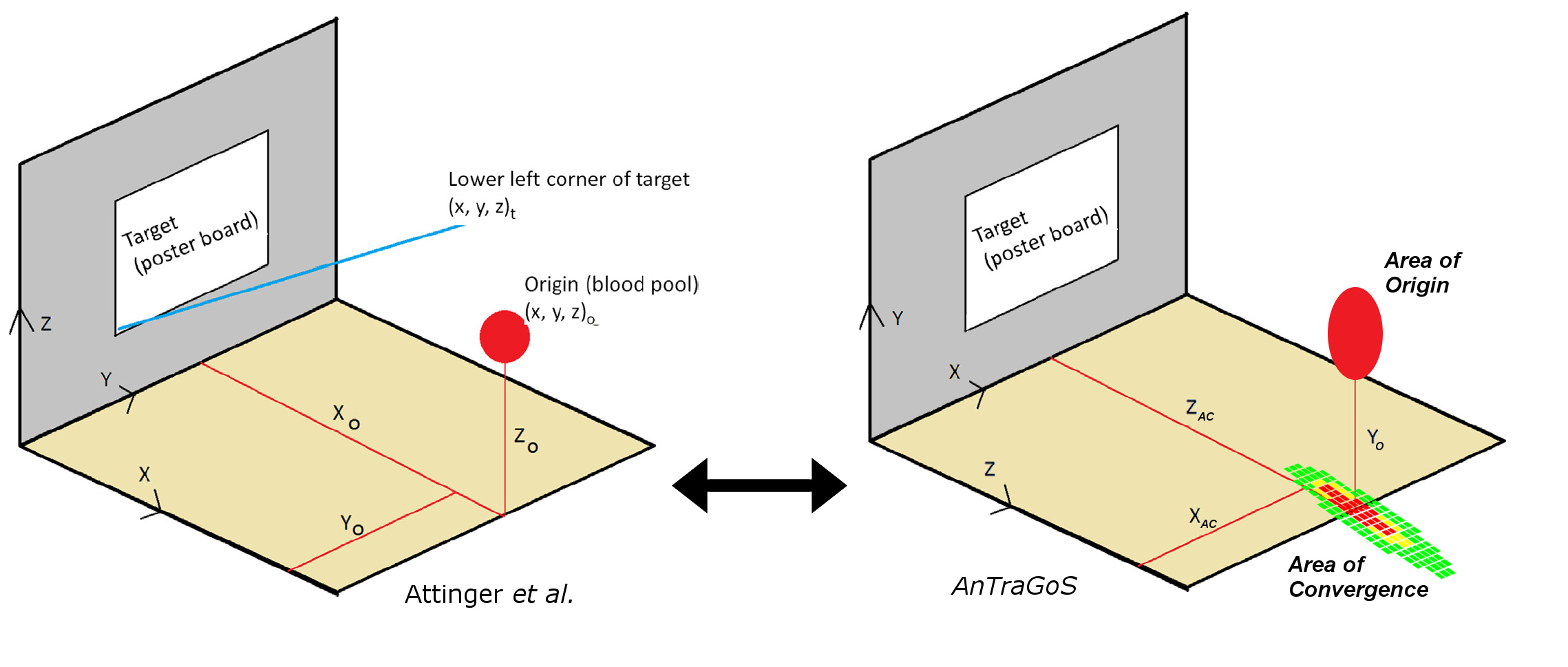}\\
\centering{Fig.1: coordinates conversion scheme}
\end{minipage}

\begin{center}
	\begin{tabular}{|c|c|c|}
		\hline
		\multicolumn{1}{|c|}{Coordinate} & \multicolumn{1}{l|}{\textbf{Attinger \textit{et al.} in \cite{A2}}} & \multicolumn{1}{l|}{\textbf{In this work and in \textit{AnTraGoS}}} \\ \hline\hline
		{ Vertical}                              & z                                             & \emph{y}                                      \\ \hline
		{ Distance from wall}                    & x                                             & \emph{z}                                      \\ \hline
		{ Horizontal coordinate}                 & y                                             & \emph{x}                                      \\ \hline
	\end{tabular}
\end{center}
\vspace{.3cm}	

\subsection{Area of Convergence}
The data in Table 1 summarize the results obtained by \emph{AnTraGoS} and the comparison between the software calculations and the real experimental setup. Details relative to single sets can be checked in the 19 attached technical sheets in \textit{Appendix 1}.
For each set of bloodstains, in separated columns, are reported:

\begin{itemize}
	\item The name of the bloodstain pattern set.
	\item The coordinates of the origin of the projection ($x_0$ and $z_0$). These data are derived from the files attached to the dataset in \cite{A2}.
	\item The coordinates ($x_{AC}$ and $z_{AC}$) of the point of maximum probability of convergence, according to the \textit{AnTraGoS} algorithm. These coordinates are automatically saved by \textit{AnTraGoS} in a log file.
	\item The differences $\Delta x=x_{AC}-x_0$ and $\Delta z=z_{AC}-z_0$, related to the offset between the calculated data and the experimental condition, splitted in the two horizontal axes.
	\item The radial distance $\sqrt{\Delta^2}=\sqrt{\Delta x^2 +\Delta z^2}$ between the point of maximum probability of convergence, according to \emph{AnTraGoS}, and the real point of origin of blood projection.
	\item The radial distance between the real point of origin of blood and the nearest coloured point of the \textit{Area of Convergence} (coloured squares are relative to areas where the probability of convergence is $\geq0.01\%	$).
\end{itemize}

\begin{center}
\setlength{\arrayrulewidth}{.5mm}
\setlength{\tabcolsep}{14pt}
\renewcommand{\arraystretch}{1.5}
\rowcolors{4}{green!5!blue!15}{green!2!blue!4}
	\begin{tabular}{|c|c|c|c|c|c|c|c|c|}
		\hline
	& \multicolumn{8}{c|}{Table 1: riepilogative data [Area of Convergence - AoC] (cm)} \\
		\hline
		Set &\multicolumn{2}{c|}{Blood Origin} & \multicolumn{2}{c|} {AoC center} &\multicolumn{2}{c|} {Differences} & \multicolumn{2}{c|} {Distances}\\
		Name &$x_0$ & $z_0$ &$x_{AC}$ &$z_{AC}$ & $\Delta x$ & $\Delta z$ & $\sqrt{\Delta^2}$ & D\\
		\hline
		C1 & 98 & 34 & 93 & 28 & -5 &  -6 &  8 &  2 \\
		C2 & 98 & 34 & 96 & 28 & -2 &  -6 &  6 &  1 \\
		C3 & 98 & 34 & 97 & 27 & -1 &  -7 &  7 &  0 \\
		C4 & 97 & 34 & 94 & 26 & -4 &  -8 &  9 &  4 \\
		C5 & 97 & 64 & 94 & 53 & -3 &  -11 &  11 &  -1 \\
		C6 & 97 & 64 & 96 & 49 & -1 &  -15 &  15 &  3 \\
		C7 & 97 & 64 & 95 & 48 & -2 &  -16 &  16 &  1 \\
		C8 & 97 & 64 & 93 & 52 & -4 &  -12 &  13 &  0 \\
		C9 & 96 & 124 & 97 & 168 & 1 &  44 &  44 &  -5 \\
		C10 & 96 & 124 & 90 & 141 & -6 &  17 &  18 &  -2 \\
		C11 & 96 & 124 & 98 & 153 & 2 &  29 &  29 &  -4 \\
		C12 & 96 & 124 & 95 & 120 & -1 &  -4 &  4 &  -3 \\
		HP0 & 103 & 34 & 100& 27& -3 & -7& 7& 1 \\
		HP10 & 99 &124 &106& 128& 7 & 4& 8&-3 \\
		HP30 & 100 &65 &96& 63& -4 & -2& 4&0 \\
		HP50 & 96 &64 &97& 60& 1 & -4& 4&-5 \\
		HP60 & 97 & 64 & 93 & 55 & -4 & -9 & 10 & 0 \\
		HP63a & 97 &124 &103& 119& 6 & -5& 8&-3 \\
		HP63b & 56 &124 &57& 85& 1 & -39& 39&-5 \\
		\hline
		\hline
		\multicolumn{2}{|c}{ } &\multicolumn{1}{c}{ } &\multicolumn{2}{c|} {\textbf{AVG {\scriptsize$\pm\Delta$}}} &  -1 {\scriptsize $\pm 3$}&-3 {\scriptsize $\pm 15$}&14 {\scriptsize$\pm 12$}&-1 {\scriptsize$\pm 3$} \\
		\hline
	\end{tabular}

\end{center}

\vspace{.3cm}	
The differences $\Delta x$ are extremely low and, after averaging over the all the sets, they balance each other extremely well: the average of these differences is -1, with a standard deviation of $\pm$ 3, which is in accordance with the symmetry of the pattern (the origin is almost centered with the images, along the horizontal axis and so, in principle we don't expect a left or right bias in the results).

The differences $\Delta z$ are instead larger and even if - on average - they balance again pretty well (the average of these differences is -3), the standard deviation is five times larger than that in the $x$ direction (standard deviation is $\pm$ 15) and there is a clear difference between the behaviour at short distance (for ranges up to 1 m there is always an underestimation of the coordinate $z_{AC}$) and the results obtained for larger distances (where the areas of convergence are also much wider and elongated).

By considering the $x$ and $z$ coordinates together, the result is that the average distance between the blood origin and the center of the area of convergence is 14 cm (with a standard deviation of 12 cm, mostly due to the contribution of the uncertainty in the $z$ axis). Considering the different experimentals conditions for the analyzed sets, one may derive from the Table 1 that the absolute offset between the two points - along the $z$ axis - is on average 1.7 cm every 10 cm of distance from the impact surface (standard deviation $=1.0$ cm), while the analogous discrepancy along the $x$ axis is only 0.3 cm (standard deviation $=0.2$ cm).

The different behaviour along the two axes is expected and well known \cite{BSL}, and essentially due both to the mechanism of projection of the drops, which do not originate from a single point in space but rather depart from different points of the borders of a flying lamina of projected blood \cite{BHC,KSW}, and to the geometrical configuration of the angles of impact (which imply, for the same angular uncertainty, larger uncertainties along the $z$ axis and smaller errors along its perpendicular direction \cite{CIF,R}).

\subsection{Area of Origin}\label{areaOrigin}
The determination of the height of origin of projections is really a problematic issue in BPA: it is the most complex analysis performed in computational BPA and the one whith the larger errors in the output. 

The stringing method assumes straight trajectories, which always and naturally generates a systematic overestimation of the vertical quota. This discrepancy increases with increasing distances between the origin and the impact surface. That is why \textit{AnTraGoS} does not make use of this method to determine the value of the height of projection.

In the field of the area of origin determination, the ideas described in \cite{A1} are certainly the most innovative and scientifically based; however the results are again affected by large errors (in this case: statistical and non systematic errors). In this work we do not exaxctly follow the procedure of \cite{A1}, but we profit of the \textit{AnTraGoS} algorithm for the numerical solution of the equations of motion to make a statistical analysis of the velocities of the drop, at a probe-point above the area of convergence. The heights which result to be a minimum for the averages and the standard deviations of the speed of the drops are argued to be the heights around which the real, experimental origin is located. As a side effect of this choice, by comparing the real and the estimated heights of origin, we also check the validity of this assumption.

For details relative to single sets please refer again to the 19 attached technical sheets in \textit{Appendix 1}. Table 2 summarizes the results. For each set of bloodstains, in separated columns, are reported:

\begin{itemize}
	\item The name of the bloodstain pattern set.
	\item The coordinate $y_0$ of the origin of the projection. This datum is derived from the files attached to the dataset in \cite{A2}.
	\item The range proposed by \textit{AnTraGoS} for the minimum and maximum value of the height of origin. This interval is identified by the values of the \textit{y}-coordinate which minimize the average and/or the standard deviation of the speed of the drops at the point of origin (from now on: \textit{best value}), including heights which present a 10$\%$ deviation from these minima. Possible data in parentheses are relative to the same calculation, but performed excluding the drops whose height - calculated with the stringing method - result too close to the value of $y_0$ (within 5 cm).
	\item The distance D between $y_0$ and the upper level of the range indicated in the previous column. For the data in parentheses see above.
	\item The distance d between $y_0$ and the \textit{best value}. For the data in parentheses see above.
\end{itemize}
The last line of the table presents the averages of the data of the corresponding column and the standard deviation.

\begin{center}
\setlength{\arrayrulewidth}{.5mm}
\setlength{\tabcolsep}{14pt}
\renewcommand{\arraystretch}{1.5}
\rowcolors{4}{green!5!red!15}{green!2!red!4}
\begin{tabular}{|c|c|c|c|c|c|}
	\hline
	& \multicolumn{5}{c|}{Table 2: riepilogative data [height of origin] (cm)} \\
	\hline
	Set &Blood Origin & \multicolumn{2}{c|} {Proposed range} & {Min distance} & {Distance}\\
	Name & $y_0$ & \multicolumn{2}{c|} {$y_{min}\triangleright y_{max}$} & D &  d\\
	\hline
	C1 & 82 &\multicolumn{2}{c|} {$60\triangleright78$ ($64\triangleright84$)}  &  4 (-2) &  10 (6) \\
	C2 & 82 &\multicolumn{2}{c|} {$68\triangleright82$ ($62\triangleright86$)}  &  0 (-4) &  6 (4) \\
	C3 & 82 &\multicolumn{2}{c|} {$58\triangleright82$ ($60\triangleright80$)}  &  0 (2) &  8 (10) \\
	C4 & 82 &\multicolumn{2}{c|} {$54\triangleright80$ ($50\triangleright84$)}  &  2 (-2) &  10 (13) \\
	C5 & 83 &\multicolumn{2}{c|} {$62\triangleright76$ ($60\triangleright81$)}  &  7 (2) &  14 (10) \\
	C6 & 83 &\multicolumn{2}{c|} {$66\triangleright90$}  &  -8 &  1 \\
	C7 & 83 &\multicolumn{2}{c|} {$50\triangleright80$ ($52\triangleright84$)}  &  3 (-1) &  13 (11) \\
	C8 & 83 &\multicolumn{2}{c|} {$54\triangleright78$ ($52\triangleright80$)}  &  5 (3) &  15 (12) \\
	C9 & 83 &\multicolumn{2}{c|} {$54\triangleright88$}  &  -5 &  5 \\
	C10 & 83 &\multicolumn{2}{c|} {$70\triangleright102$}  &  -19 &  -12 \\
	C11 & 83 &\multicolumn{2}{c|} {$20\triangleright58$ ($50\triangleright78$)}  &  25 (5) &  45 (19) \\
	C12 & 83 &\multicolumn{2}{c|} {$58\triangleright72$ ($50\triangleright80$)}  &  11 (3) &  19 (11) \\
	HP0 & 82 &\multicolumn{2}{c|} {$60\triangleright78$ ($56\triangleright86$)}  &  4 (-4) &  12 (6) \\
	HP10 & 83 &\multicolumn{2}{c|} {$60\triangleright96$}  &  -14 &  3 \\
	HP30 & 83 &\multicolumn{2}{c|} {$40\triangleright64$ ($50\triangleright78$)}  &  19 (5) &  31 (21) \\
	HP50 & 83 &\multicolumn{2}{c|} {$70\triangleright108$}  & -25 &  -13 \\
	HP60 & 83 &\multicolumn{2}{c|} {$50\triangleright80$}  & 3 &  18 \\
	HP63a & 82 &\multicolumn{2}{c|} {$20\triangleright60$ ($50\triangleright86$)}  &  23 (-3) &  38 (9) \\
	HP63b & 82 &\multicolumn{2}{c|} {$70\triangleright108$}  &  -25 &  -13 \\
	\hline
	\hline
	\multicolumn{2}{|c}{ } &\multicolumn{2}{c|}  {\textbf{AVG {\scriptsize$\pm\Delta$}}} & 1 {\scriptsize$\pm14$} (0 {\scriptsize$\pm3$})&11 {\scriptsize$\pm16$} (11 {\scriptsize$\pm5$}) \\
	\hline
\end{tabular}
\end{center}
\vspace{.3cm}	

Some comments about the results reported in Table 2 are needed.
First of all we notice that all the projections in the dataset have been produced at the same height of about 82--83 cm. \textit{AnTraGoS} often proposes a range which is lower than this value, both including all the stains and excluding the lower trajectories. In 16 cases (against 3) the \textit{best value} suggested by the software is lower than the experimental value: the average \textit{distance} between these values is 11 cm, with a standard deviation of 16 cm. The standard deviation is reduced to 5 cm if we exclude the lower trajectories: this suggests that excluding some trajectories (about 10 \% of the total) may improve the capability of the software to converge toward more correct results.

The ranges proposed by the software are not symmetrical, in general, with respect to the \textit{best value}. These intervals are from 12 to 40 cm wide, and the upper level of these intervals is often surprisingly close to the real experimental value of the origin. The average \textit{distance} between these points is 1 cm (0 cm if we neglect some of the stains); again the standard deviation is significantly smaller for the set where about 10 \% of the stains are discarded from the analysis (3 cm against 14 cm). The idea of excluding some of the stains from the analysis has also the following physical interpretation. The assumpion of spherical shape for the drops neglects the natural oscillation of the fluid during the flight \cite{RSL}. Nevertheless, the oscillation causes misinterpretation of the angle of impact, because in this case Eqs.\ref{eq_ang_impatto} and \ref{eq_ang_gamma} are basically no longer valid. The angle of impact may then result in a significantly lower height of origin, since many possible trajectories, above the tangent of impact, are excluded from the analysis. And this error is not statistically balanced by a symmetrical overestimation of the height due to other trajectories (as in the case of the calculation of the area of convergence), because - if we consider gravity - there is no vertical symmetry at all. Excluding some stains is then a possible mean to discard a possible source of systematic error from the analysis of the pattern.

However, it is generally true that the discrepancy between real and calculated values is increasing with the distance from the wall, and that the results are more stable and homogeneous at short distances. At short distance, the process of discarding some stains from the analysis of the height of origin is certainly to be preferred, and this is possibly due to the predominancy of the oscillating effect, in the proximity of the point of origin (see again \cite{RSL}).

\section{Discussion}\label{conclusions}
The analysis of the data of the single experimental configurations and those of the riepilogative tables of the previous paragraph suggest the following conclusions:
\begin{itemize}
	\item \textit{AnTraGoS} was able to indicate the \textit{Area of Convergence} with a very reduced error in both horizontal coordinates: along the \textit{z} axis (perpendicular to the wall) the discrepancy is on average 1.7 cm every 10 cm of distance, while along the \textit{x} axis the same discrepancy is only 0.3 cm. Considering the uncertainty (standard deviations), the conclusive result is that the expected error is 1-5\% along the \textit{x} axis and 7-27\% along the \textit{z} axis. These results are not systematic, \textit{i.e.} software dependent, but have a strong dependence on the statistical approach and on the complexity and the variables involved in the analytical process. In cases where the stains of the pattern lie on different, perpendicular walls, it can be comfortably argued by symmetry that the errors along the \textit{x} and \textit{z} directions are both equal to 1-5\%;
	\item In 13  examined cases, the real, experimental point of origin lies within the spotted area of convergence, in the other 6 cases the distance between the point and the area is 1--4 cm. The results are perfectly centered along the \textit{x} axis, while the largest mismatches are again found along the \textit{z} axis. The area of convergence calculated and drawn by \textit{AnTraGoS} is then strongly indicative of the location of the origin of the projections: it is therefore not only an area of probablity of convergence but also the figurative representation of the horizontal projection of the area where the impact spatter effectively occurred;
	\item For what concerns the height of origin and the identification of the \textit{Area of Origin}, the \textit{best value} suggested by \textit{AnTraGoS} is close to the real, experimental value, but often overestimated. Compared to values of 82--83 cm of $y_0$, the best values suggested by the software are spread in the range 38--96 cm, with an average of 72 cm. More centered results can be however obtained if the upper bound of the range proposed by \textit{AnTraGoS} is considered, instead of the \textit{best value}. Uncertainties are also larger along the \textit{y} axis, compared to those along horizontal coordinates (which is however an expected and known result \cite{A1}). Considering the different values of distance from the wall, the conclusive result is that the expected error (uncertainty) of the height of origin estimated by \textit{AnTraGoS} is - on average - 2 cm every 10 cm of distance. 
	\item \textit{AnTraGoS} performed the calculations for the determination of the area of convergence and the estimation of the height of origin in a very short time (in less than 1 s and less than 1 min, respectively), which is much less than the time needed for the stain selection and for the measurement of the stains within an image analysis software;
	\item no particular difference in the results and in the response of the software could be documented between the \textbf{C{\scriptsize \#}} and the \textbf{HP\_{\scriptsize \#}} patterns, suggesting that the software and the analytical method can be invariably used for different spatter sources;
	\item the overall analysis performed in \textit{AnTraGoS} confirms - once again - the possibility of consideration of stains pointing downwards.
\end{itemize}
In conclusion, \textit{AnTraGoS} can be used in BPA as a valid support to the identification of the area of convergence and the height of origin. The calculations and the outcomes of the algorithms have been checked both directly and by comparison with real, experimental data. The results are affected by statistical, non systematic errors, which are however within the limits of the general requirements of the forensic applications.
\vspace{1cm}
\section*{Acknowledgements}
We acknowledge the Italian \emph{Polizia di Stato} for supporting \textit{AnTraGoS} and the research projects in Bloodstain Pattern Analysis.
\newpage
\bibliographystyle{abbrv}

\renewcommand{\theequation}{A-\arabic{equation}}    
  \setcounter{equation}{0}  
  
  \setlength{\extrarowheight}{-2pt}
\section*{Appendix 1: technical sheets}\label{Appx}
In the following pages the analytical results of the overall validation process are presented, each page reporting a table with the data relative to a single data set (one experimental setup per page). Every summary table is divided into four blocks: two in the first row, and two in subsequent, separate rows. All the data are in centimeter units, rounded up to the nearest integer value. According to the \textit{AnTraGoS} unit system, the vertical line is aligned with the \textit{y} axis (upward oriented), while the horizontal plane is the \textit{xz} plane, with\textit{ x} aligned with the deposit surface. \newline
\begin{itemize}
	\item The first row is divided into two columns:
	\begin{itemize}
		\item In the first column a table summarizes the \textbf{coordinates of the stains} which have been selected for the analysis. Every stain has a name composed of the name of the set and a progressive index (\textit{e.g.} the stain C1\_3 is the \textit{3rd} selected stain in the \textit{C1} data set).
		\item The second column shows a \textbf{sketch of the \textit{Area of Convergence}}, as calculated by \textit{AnTraGoS}. It is clearly a top view, with the scale indicated at the bottom left corner. The colors of the area of convergence are indicative of the different values of probability of convergence in every single square:\newline \textbf{red}: $Prob>1.5\%$; \textbf{yellow}: $0.5<Prob\leq 1.5\%$; \textbf{green}: $0.01\%<Prob\leq 0.5\%$.
	\end{itemize}
	\item The second row reports the \textbf{analytical results of the \textit{Area of Convergence}}. The first table is relative to the \textbf{\textit{x}-coordinate}, and the five columns, from left to right, show the values of: 1- the coordinate of maximum probability of convergence (calculated via \textit{AnTraGoS}); 2- the range of coordinates of the coloured squares (points calculated via \textit{AnTraGoS} with $Prob\geq 0.01\%$); 3- the experimental value of the coordinate of the point of origin (datum derived from \cite{A2}); 4- the distance along the axis between the points in column 1 and 3; 5- the distance of the point in column 3 with the interval indicated in column 2 (a negative sign means that the point is located inside the interval).\newline The second table reports the same data as the first, but relative to the \textbf{\textit{z}-coordinate}.\newline Finally, outside the tables, the (diagonal) distance between the points ($x_{AC}$,$z_{AC}$) and ($x_0$,$z_0$) can be found, together with the indicative distance of the point ($x_0$,$z_0$) from the coloured area (a negative value means that the point is within the area itself).
	\item The third and last row presents a \textbf{schematic overview concerning the \textit{Height of origin}}. The value $y_0$ is again derived  from \cite{A2}. The graphs represent some indications useful to investigate the height of origin of the projection: on the left side, next to a 120 cm leveling rod, there are some circles indicating the \textit{y}-coordinate above the point of maximum probability of convergence, referred to the tangent lines to the trajectories at the point of impact of the stains (tangent method, straight lines). No uncertainty is shown in this case, for clarity reasons. On the right, next to another leveling rod, three differently coloured histograms are depicted (vertical spacing: 2 cm): the \textbf{black} one is relative to the values of the \textit{average initial speed} of the drops, as calculated by \textit{AnTraGoS} for that particular \textit{y}-coordinate; the \textbf{white} one is relative to the \textit{standard deviation of the initial speed} (the standard deviation of the values of speed calculated for all the drops, at that particular \textit{y}-coordinate); the \textbf{red} one is relative to the \textit{maximum speed} (speed of the fastest projected drop in the group). For every coloured histogram, the difference of length of the bars is proportional to the difference of the corresponding calculated values. The red arrows indicate the value of $y_0$.\newline On the right, next to the graph relative to all the selected stains, another graph can be present: this second graph is built in the same way, but it is relative to a reduced set of stains, discarding all the drops whose height calculated with the stringing/tangent method result too close to the value of $y_0$. These drops interfere too much with the calculation of the original speed, because they are too close to the calculated tangents of impact and therefore they result to be too fast: referring to the considerations of Par. \ref{areaOrigin}, they are regarded as statistical outliers.\newline Below the graph(s) three lines of text summarize the following data: 1- the range proposed by \textit{AnTraGoS} refers to the minima of the \textbf{black} and \textbf{white} histograms shown above (including bars with values within a 10$\%$ from the minima); 2- the distance of $y_0$ from range indicates the distance of the coordinate $y_0$ from the upper value of the interval in line 1; 3- the distance of $y_0$ from \textit{best value} indicates the distance of the coordinate $y_0$ from the average of the minima of the \textbf{black} and \textbf{white} histograms. Repeated tests performed by the authors have shown that the minimizations of averages and standard deviations of original speed of drops can be strongly indicative of the height of origin of projection and can be used to reduce the statistical range of uncertainties described in \cite{A1}.
\end{itemize}
\newpage
\begin{center}
	\textbf{SetC1: cylinder rig}
\end{center}
\rowcolors{15}{green!5!blue!15}{green!2!blue!4}
\centering
\renewcommand{\arraystretch}{1.5}
\setlength{\arrayrulewidth}{.5pt}

	\newpage\begin{center}
		\textbf{SetHP\_0: hockey puck rig}
	\end{center}
	\rowcolors{15}{green!5!blue!15}{green!2!blue!4}
	\centering
	\renewcommand{\arraystretch}{1.5}
	\setlength{\arrayrulewidth}{.5pt}
	%
	\newpage\begin{center}
		\textbf{SetHP\_10: hockey puck rig}
	\end{center}
	\rowcolors{15}{green!5!blue!15}{green!2!blue!4}
	\centering
	\renewcommand{\arraystretch}{1.5}
	\setlength{\arrayrulewidth}{.5pt}
	%
	\newpage\begin{center}
		\textbf{SetHP\_30: hockey puck rig}
	\end{center}
	\rowcolors{15}{green!5!blue!15}{green!2!blue!4}
	\centering
	\renewcommand{\arraystretch}{1.5}
	\setlength{\arrayrulewidth}{.5pt}
	%
	\newpage\begin{center}
		\textbf{SetHP\_63a: hockey puck rig}
	\end{center}
	\rowcolors{15}{green!5!blue!15}{green!2!blue!4}
	\centering
	\renewcommand{\arraystretch}{1.5}
	\setlength{\arrayrulewidth}{.5pt}
	%
	\newpage\begin{center}
		\textbf{SetHP\_63b: hockey puck rig}
	\end{center}
	\rowcolors{15}{green!5!blue!15}{green!2!blue!4}
	\centering
	\renewcommand{\arraystretch}{1.5}
	\setlength{\arrayrulewidth}{.5pt}
	%
}\\ \hline 
	\end{tabular}
	\newpage

\bibliographystyle{abbrv}


\end{document}